\documentclass{PoS}

\newcommand{\emer}{e-MERLIN}
\newcommand{\lovt}{Lovell Telescope}
\newcommand{\lov}{LT}
\newcommand{\goon}{Goonhilly}

\newcommand{\aap}{A\&A}

\newcommand{\mnras}{MNRAS}

\title{Goonhilly: a new site for e-MERLIN and the EVN}

\ShortTitle{Goonhilly: a new site for e-MERLIN and the EVN}

  \author{\speaker{H.-R. Kl\"{o}ckner}$^{1,2}$, S. Rawlings$^{1}$, I. Heywood$^{1}$ , R. Beswick$^{3,4}$, T. W. B. Muxlow$^{3}$, S. T. Garrington$^{3,4}$, J. Hatchell$^{5}$, M. G. Hoare$^{6}$, M. J. Jarvis$^{7}$, I. Jones$^{8}$, H. J. van Langevelde$^{9,10}$\\
       $^{1}$Subdepartment of Astrophysics, University of Oxford, Denys-Wilkinson Building, Keble Road, Oxford, OX1 3RH, UK\\
               $^{2}$Max-Planck-Institut f\"{u}r Radioastronomie, Auf dem H\"{u}gel 69, 53121 Bonn, Germany\\
               $^{3}$e-MERLIN / VLBI National Radio Astronomy Facility, Jodrell Bank Observatory, The University of Manchester, Macclesfield, Cheshire, SK11 9DL, UK\\
               $^{4}$Jodrell Bank Centre for Astrophysics, School of Physics and Astronomy, The University of Manchester, Oxford Road, Manchester, M13 9PL, UK\\
               $^{5}$Astrophysics Group, CEMPS, University of Exeter, Stocker Road, Exeter, EX4 4QL, UK \\
               $^{6}$School of Physics and Astronomy, University of Leeds, Leeds, LS2 9JT, UK\\
               $^{7}$Centre for Astrophysics Research, STRI, University of Hertfordshire, Hatfield, AL10 9AB, UK\\
               $^{8}$Goonhilly Earth Station Ltd, Goonhilly Downs, Helston, Cornwall, TR12 6LQ, UK\\
               $^{9}$Joint Institute for VLBI in Europe, Postbus 2, 7990 AA Dwingeloo, The Netherlands\\
               $^{10}$Sterrewacht Leiden, Leiden University, Postbus 9513, 2300 RA Leiden, The Netherlands\\
            }

            \abstract{The benefits for the \emer\ and EVN arrays of
              using antennae at the satellite communication station at
              Goonhilly in Cornwall are discussed. The location of
              this site - new to astronomy - will provide an almost
              equal distribution of long baselines in the east-west-
              and north-south directions, and opens up the possibility
              to get significantly improved observations of equatorial
              fields with \emer . These additional baselines will
              improve the sensitivity on a set of critical spatial
              scales and will increase the angular resolution of
              \emer\ by a factor of two. \emer\ observations,
              including many allocated under the e-MERLIN Legacy
              programme, will benefit from the enhanced angular
              resolution and imaging capability especially for sources
              close to or below the celestial equator (where ESO
              facilities such as ALMA will operate) of including the
              Goonhilly telescopes. Furthermore, the baselines formed
              between Goonhilly and the existing stations will close
              the gap between the baselines of \emer\ and those of the
              European VLBI Network (EVN) and therefore enhance the
              legacy value of \emer\ datasets.  }

\FullConference{10th European VLBI Network Symposium and EVN Users Meeting: VLBI and the new generation of radio arrays\\
		September 20-24, 2010\\
		Manchester Uk}

\begin{document}

\section{Introduction}

\noindent Half a century ago, visionary engineers and scientists
designed and built the world's first satellite communication station
choosing \goon\ in Cornwall as its location (lat=50.0504$\,^{\circ}$
North, lon = 5.1835$\,^{\circ}$ West). Since then, \goon\ has
pioneered many of the advances in global communication and inspired
budding scientists and engineers. Now in 2011, most needs for
satellite and tele-communications are met by using a combination of smaller antennae and
undersea cables, and there are plans afoot to use Goonhilly's valuable
assets for a variety of new purposes including radio astronomy.
\noindent There are three 30-m-class telescopes at \goon, and two of these
could straightforwardly be adapted for radio astronomy. Goonhilly-1
(or "Arthur") can be most easily incorporated into \emer\ as its
design was based on the Jodrell Bank MKII Telescope that is already a
mainstay of the network. Goonhilly-3 (or "Guinevere") is perhaps a more suitable
Goonhilly antenna to use for higher-frequency (especially K-band)
observations.

Assuming that the chosen antennae can be outfitted with standard
\emer\ L-, C- or K-band receivers, and connected to the eighth "spare"
input of the correlator (information on \emer\ can be found via
http://www.e-merlin.ac.uk), the improvements that the addition of a
\goon\ antenna would bring to the enhanced array performance, and
consequently the scientific return of the \emer\ Legacy observing
programmes, are discussed.

\section{Including  Goonhilly and using the Lovell Telescope in
  the \emer\ array}

\noindent The imaging capability of an interferometer is related to
the distribution of its antennae and their individual
sensitivities. Here the UV plane and imaging benefits are discussed if
a "Goonhilly" antenna\footnote{Note that in addition to the
  already mentioned telescopes at the Goonhilly side there is another
  30-m-class telescope (Goonhilly 6) available. If all three 30-m-class telescopes could be
  phased up then the equivalent of a 50-m-class telescope could be delivered by Goonhilly resources.}
is included in the \emer\ array and if the \lovt\ (\lov) is used.

\subsection{UV plane benefits}

\begin{figure}
\begin{center}
\includegraphics[width=.45\textwidth]{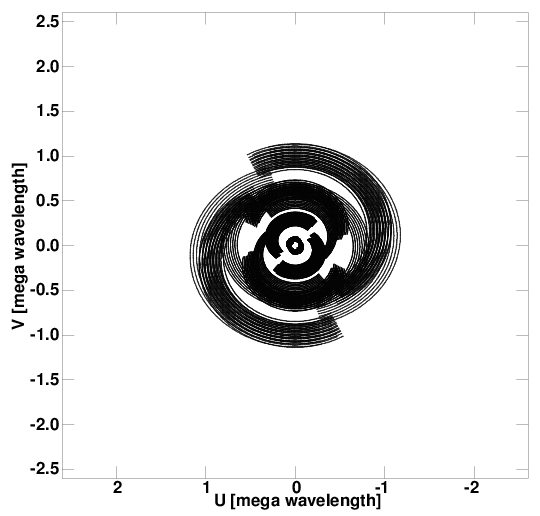}
\includegraphics[width=.45\textwidth]{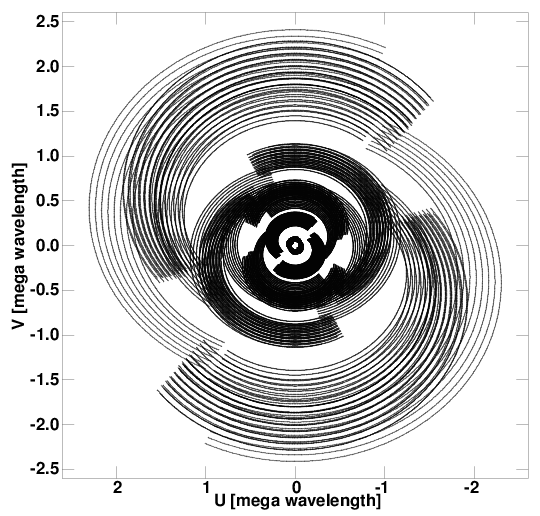}\\
\includegraphics[width=.45\textwidth]{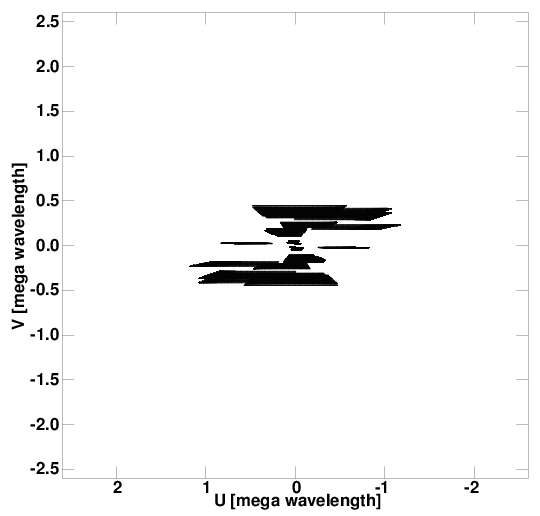}
\includegraphics[width=.45\textwidth]{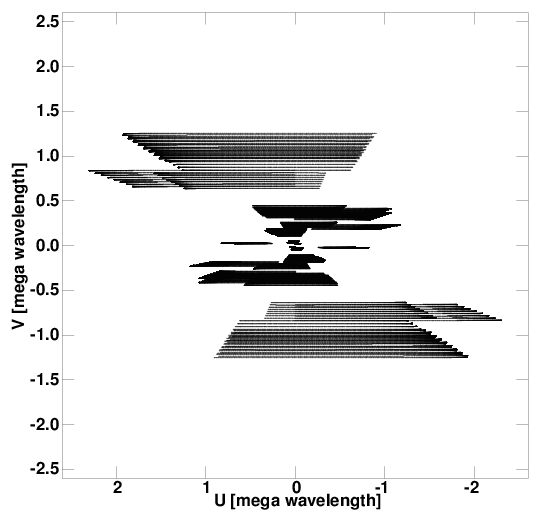}\\
\end{center}
\caption{The left column shows the UV coverage of the \emer\ array consisting of the Lovell Telescope [76~m]
  (the MKII [25~m] was not used), Cambridge [32~m], Defford [25~m], Knockin
  [25~m], Darnhall [25~m] and Pickmere [25~m]) telescopes. The right column shows UV coverages generated by including
  the Goonhilly telescope. The UV coverages in the top row are
  based on a 12-hour, L-band simulation of a source at a declination
  of 60$\,^{\circ}$, whereas the UV coverages in the lower row are for a
  source at declination 0$\,^{\circ}$. Note that the UV coverage
  exhibits a radial spread as a function of frequency, and the
  individual tracks of the baseline should thus practically fill the
  UV plane when \emer\ observes in full multi-frequency synthesis
  mode. For clarity however, the simulation here presents a processed
  bandwidth of 400~MHz which has been split into 8 channels of 50~MHz
  each and therefore individual tracks are clearly visible,
  particularly for the longest baselines. }
\label{fig1}
\end{figure}

\noindent Generally the angular resolution of an interferometer is determined by
the longest "projected" baseline (the value of which [in radians] is
approximately $\lambda / b$, for which $\lambda$ is the
observing wavelengh and b is the length of the longest baseline,
see e.g. Rohlfs \& Wilson 2004).  With its current setup, \emer\ has a
maximum baseline of 217~km which provides an angular resolution
of 170~milli-arcsec (mas) at L-band. Introducing one of the
telescopes at Goonhilly would increase the maximum baseline to
440~km, thus doubling the angular resolution of e-MERLIN,
bringing it to values of 82~mas, 17~mas,
and 6~mas at L-, C- and K-band respectively. 

Figure~\ref{fig1} affords a more detailed view on the
improved imaging capabilities, showing the L-Band UV coverages of
\emer\ and \emer\ including Goonhilly at two different declinations
(60$\,^{\circ}$ and 0$\,^{\circ}$). The UV coverage is based on a
12-hour synthesis\footnote{The UV coverages are derived from an array
  simulation using the AIPS task {\tt UVCON}.}. For L-band
observations the \emer\ bandwidth will be of the order of 400~MHz
(1.3--1.7 GHz), which is simulated as 8 channels of 50~MHz bandwidth
each. At 60$\,^{\circ}$ declination the UV coverage is circularly
symmetric and would almost fully cover all spatial frequencies up to
the longest baselines. Due to the location of Goonhilly the array
would have almost equally long east-west and north-south baselines,
symmetrically extending the UV-coverage by a factor of $\sim$2. Such a
configuration would significantly improve the snap-shot imaging
capability with respect to the current \emer\ configuration.  As the
declination of an observation moves towards zero the UV coverage
becomes increasingly "squashed" in the north-south direction, which results in an elongated
synthesized beam\footnote{The UV coverage and synthesised beams are
  directly coupled in that they are Fourier transforms of one
  another.}. This is a direct result of having "projected" baselines
which directly influences the imaging capability of the \emer\
interferometer. For example, a 12-hour observation on the celestial
equator and using uniform weighting would result in angular resolutions of 320~mas$\times$140~mas
for e-MERLIN, and 140~mas$\times$72~mas with the additional station at
Goonhilly. For further discussion on equatorial imaging using \emer\
and telescopes located in Chilbolton and in Goonhilly see Heywood et al. (2008,
2011).

In order to fully exploit such UV coverage new wide-band
imaging algorithms must be used. Properly accounting for sources which
are morphologically complex and exhibit components with different
spectral indices is a major challenge for calibration and
imaging. This is not the case for spectral line observations during which
the UV coverage advantage gained by multi-frequency synthesis is lost,
as each spectral channel will only occupy a narrow region of the UV
plane.

\subsection{The image sensitivity}

\noindent Adding a new antenna into an existing array provides
enhanced imaging sensitivity and depending on its location it will
potentially enhance the spatial sensitivity by adding baselines that
traces new sets of spatial scales. Here some basic radio astronomy
relationships are presented in order to quantify such benefits.

The baseline sensitivity [in units of Jy] at a single polarisation in the real or the imaginary component of the visibility 
formed between antennas $i$ and $j$ is
given by:
\begin{eqnarray}
\Delta S^2_{ij} = \eta^2_{b} \, \frac{T_{i} T_{j}}{2 \Delta t \Delta \nu K_{i} K_{j}}, 
\end{eqnarray}

\noindent where $\eta^2_{b}$ is an efficiency term accounting for
losses due to digitisation of the signal for correlation, $T_{\,i}$ is
the system temperature of antenna $i$ or $j$ respectively [K], and for
example $K_{\,i}$ is the antenna sensitivity (where $K_{\,i} =
(\eta_{i} A_{i})/(2 k)$; $\eta_{i}$ is the antenna efficiency, $A_{i}$
is the geometrical area [m$^2$], and $k$ is the Boltzmann constant),
$\Delta t$ is the integration time per visibility [s], and $\Delta
\nu$ is the bandwidth [Hz]. In order to detect a single source, and
therefore to calibrate or self-calibrate the data, a sensitivity of
5~$\times$~$\Delta S_{\, ij}$ is required within the coherence time
(Walker 1989). A practical measure of the coherence time is the
averaging time at which the scalar- and the vector-averages of the
phases or the amplitudes are significantly different (e.g. the
amplitude vector average is calculated by $\sqrt{<re>^2 + <im>^2}$ and
the amplitude scalar average via $<\sqrt{re^2+im^2}>$, where $<>$
indicate the time averages and $re$ and $im$ are the real and imaginary
component of the visibilities).

The imaging sensitivity $\Delta I$ [Jy] is closely related to the
baseline sensitivity and for an inhomogeneous array can be determined
via:
\begin{eqnarray}
{\Delta I} = \frac{2 k \eta_{b}}{\sqrt{N_{\rm Stokes} \Delta \nu \Delta t}} \,\, \frac{1}{\sqrt{ (\sum \frac{\eta_i A_i}{T_i})^2 - \sum (\frac{\eta_i A_i}{T_i})^2 }},
\end{eqnarray}

\noindent where $N_{\rm Stokes}$ is the number of polarisation
products and $\Delta t$ is now the total on-source time. In case of a homogeneous array the image sensitivity is given by: 
${\Delta I} = (\eta_{b} SEFD) / \sqrt{N\,(N - 1) \, N_{\rm Stokes} \Delta \nu \Delta t}$, where N is the number of antennae, SEFD is the system equivalent flux density, which is defined by SEFD = $T_i /
  K_i$. For a full discussion on the derivation of the above equations see Walker (1989)
and Wrobel \& Walker (1999).

For a basic estimate of the relative improvement in imaging
sensitivity the following L-band SEFDs are considered: Lovell Telescope
36~Jy, MKII Telescope 350~Jy, and 220~Jy for the Telescope in
Cambridge (these SEFDs are provided in the EVN status table). For all other antennae in the array a SEFD of 350~Jy is
assumed. In this case the image sensitivity of \emer\ will improve by
a factor of $\sim$1.9 if the Lovell Telescope is used instead of the
MKII Telescope. Furthermore, comparing \emer\ including the Lovell
Telescope and Goonhilly with \emer\ and the MKII Telescope shows an even higher
imaging sensitivity of a factor of $\sim$2.1 or equivalently allows a reduction in the
observing time by a factor of $\sim$4.4. An equal sensitivity
improvement is expected for C-band observations, because the antennae
have similar SEFD values. For K-band observations this situation
changes significantly; the effect of adding sensitivity with the
Goonhilly Telescopes is of importance and of the order of $\sim$20\%,
as neither the Lovell nor the Defford Telescopes cannot operate at these
frequencies. Note that these sensitivity improvements are valid for
both continuum and spectral line observations.

In addition, to the imaging sensitivity is the spatial
sensitivity of the additional baselines formed by the antennae in
Goonhilly. The baselines in the triangle
Jodrell-Bank-Cambridge-Goonhilly would build the most sensitive part
of the \emer\ array and is particular important when the Lovell
Telescope is included to increase the long-baseline sensitivity.

\begin{figure}
\begin{center}
\includegraphics[width=.45\textwidth]{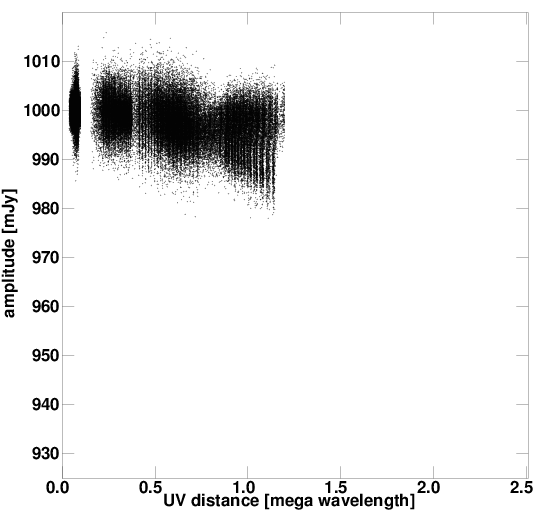}
\includegraphics[width=.45\textwidth]{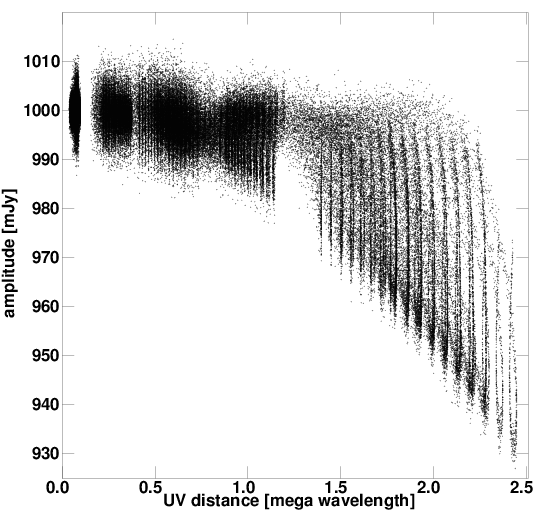}\\
\includegraphics[width=.45\textwidth]{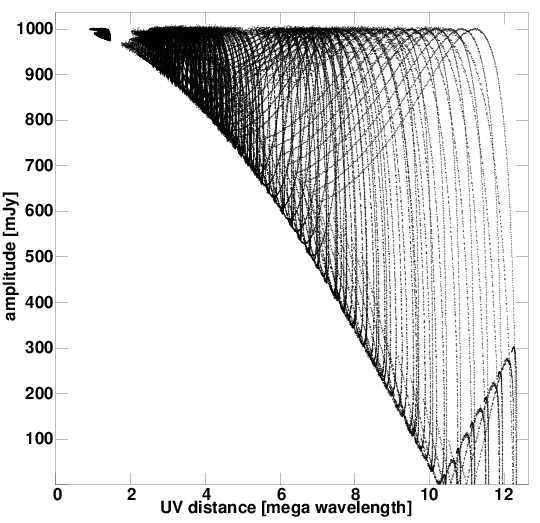}
\includegraphics[width=.45\textwidth]{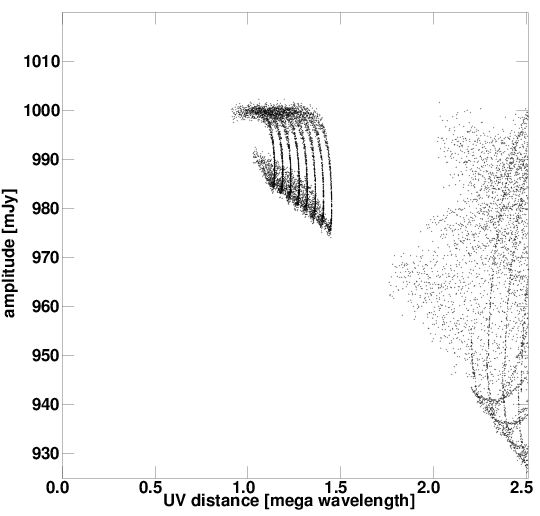}
\end{center}

\caption{Simulated amplitude versus UV distance plots for \emer\
  (upper left), \emer~+~Goonhilly (upper right) and the EVN (lower
  left and lower right).  An observation of two sources at
  60$\,^{\circ}$ declination with equal flux densities of 0.5~Jy and
  separated by 9~mas have been simulated with the AIPS tasks {\tt
    UVCON} and {\tt UVSUB}. The simulation here presents a processed
  bandwidth of 400~MHz which has been split into 8 channels of 50~MHz
  each.  The plot in the lower left displays the full EVN simulation,
  whereas the lower right shows a sub-section of the EVN simulations
  in order to highlight the shortest baselines (Effelsberg--Westerbork
  baseline covers a range of 1--1.5 mega wavelength). The EVN simulations
  are based on the following telescopes: Jodrell Bank (Lovell
  Telescope), Effelsberg, Westerbork (phased-up), Onsala, Noto, Torun, Medicina.}
\label{fig2}
\end{figure}

\subsection{\emer\ and the European VLBI Network (EVN)}

\noindent The combination of \emer\ and EVN observations provide a
unique view of the nuclear composition of extragalactic
sources. The broad range of baselines within the combined datasets
correspondingly probe a broad range of important linear
size scales in extragalactic sources, from kilo- to (sub-)parsec.

With the anticipated imaging sensitivities of \emer\ normal
galaxies, starburst galaxies, radio quiet quasars, and hybrid
AGN-starburst galaxies will be observable at high redshifts and with mas
resolution. It is of vital importance to disentangle the
nuclear composition of these galaxies e.g. in order to determine the AGN
contribution to the total radio luminosity (Kl\"ockner et
al. 2009). However there is observational evidence that radio
emission can be missed between the baselines of the EVN and \emer . For
example, observations of the AGN-starburst system
J123642$+$621331 at redshift z=4.424 (Muxlow et al. 2005, Garrett et
al. 2001) show that the radio flux density does not change from
arcsecond to subarcsecond resolution between the baselines provided by the WSRT (489~$\mu$Jy),
the VLA-A (467~$\mu$Jy), and the MERLIN (472~$\mu$Jy) arrays, but
it clearly changes at the higher EVN resolution (248~$\mu$Jy). The missing-flux effect is
not only seen in continuum observations; it has also been noted for
spectral line emission, e.g. for the OH Megamaser emission in Mrk~273 (Yates et
al. 2000, Kl\"ockner \& Baan 2004).

Currently the only method available to compensate for such an effect is to perform EVN
observations which include the Lovell/MKII and the Cambridge
Telescopes from the e-MERLIN array. Obviously the time available to be allocated to such projects is rather
limited. If such an observation is not
possible, the only option one has is to combine both datasets by scaling the visibilities according to
the Effelsberg - Westerbork baseline (270~km) and the
MERLIN flux. In practice, the scaling factor is determined via the flux
density in the image plane and applied to both datasets before combination (e.g. via the AIPS
task {\tt DBCON}). Such a technique is only reliable if the source
structure is close to a point source. However if the target
source exhibits more complicated structure the reliability of such combined
MERLIN (or in future e-MERLIN) and EVN data sets is questionable.

Figure~\ref{fig2} illustrates the potential dangers of ill-defined
flux density measurement in both arrays. These amplitude versus
UV-distance plots are based on \emer\ and EVN simulations of two point
sources separated by 9~mas and with a flux density of about 0.5~Jy
each. Such a model would not be resolved by \emer\ as shown in the top
left-hand plot of Figure~\ref{fig2}, but when Goonhilly is added to
the array the source structure is detected by the additional
baselines. Furthermore, comparing the \emer\ simulation with those of
the EVN shows that there is hardly any overlap of the visibilities in
UV-distance. This spread in amplitude due to the source structure will
make it impossible to reliably model the datasets. However, including
a station at Goonhilly would allow reliable modeling and
cross-matching of the flux density at 1~M$\lambda$ to 2.5~M$\lambda$
scale sizes. Crucially, the baselines would fill the gap in the EVN
data and therefore provide new structural information\footnote{Note
  that once the focal plane array program in Westerbork (APERTIF) is
  installed (by $\sim$2012) the C-Band sensitivity of the
  Effelsberg-Westerbork baseline drops and roughly equals the
  sensitivity of the baseline between the Lovell Telescope and
  Goonhilly.}. Once APERTIF is installed at the WSRT it would be of
particular importance if one of the Goonhilly antennae would be included to the
EVN observing session. Such setup would provide an almost uniform
structural sensitivity and would be ideal for the "the world's
leading real-time VLBI array" that is compiled by the EVN telescopes
(Szomoru et al. 2011).

The location of Goonhilly seems to be ideal in order to
produce better snap-shot imaging quality than is currently possible
with \emer . Furthermore, the baselines formed between Goonhilly and
Jodrell Bank, and Goonhilly and Cambridge (440~km) provide a
unique structural probe at 120~mas, which translates to kilo-parsec
scales at redshift 1. In addition, the new baselines formed by the
Goonhilly station will ensure that data from \emer\ can at all times
be reliably combined with EVN datasets due to the overlap in baseline
lengths.  As discussed previously this will afford a totally unique
view of radio emission from galactic scales (10~km baseline Jodrell Bank to
Pickmere) to the central regions of galaxies.

\section{Impact on Legacy proposals}

\noindent The \emer\ Legacy programme consists of eleven projects
which spans a wide range of astrophysical themes and will address many
current challenges in astronomy and astrophysics. In particular, the
projects build a coherent set of surveys that addresses key questions
in star/planetary system formation, in radio outflow/jet physics, in
the interplay between accreting black holes and nuclear starbursts, and
galaxy/AGN evolution. The data products generated by the Legacy
programmes will be made available for exploitation by the whole
astronomical community.

Here a general overview of the basic observational
requirements of these programmes is given. The quoted hours are the allocated
hours for the proposals and not the proposed ones. The benefits in imaging
sensitivity of including the Lovell Telescope (LT) in the array is discussed. Furthermore, the enhanced angular
resolution brought by including a telescope at Goonhilly (GH) is
discussed. The \emer\ Legacy programmes and their
basic requirements are:

{\scriptsize
\begin{itemize}
\item  {\bf A}strophysics of {\bf GA}laxy {\bf T}ransformation and {\bf E}volution (PI's: C.~Simpson, I.~Smail):\\
{\bf -} L-Band, targeted imaging (single pointings), 330~hours.\\
{\bf -} The LT is not considered due to FoV limitations, but the increased angular resolution due to GH 
enables the study of the radio morphology at $\sim$500~pc scales (an average redshift of 0.5 is assumed).

\item The \emer\ {\bf C}yg {\bf OB}2 {\bf RA}dio {\bf S}urvey: Massive and Young stars in the Galaxy (PI: R.~Prinja):\\
{\bf -} L-Band, C-Band, several pointings to mosaic, multiple epochs at C-band, 294~hours.\\
{\bf -} Using the LT allows the detection of some 50 O stars, whereas without it only the most massive 
stars can be observed, thus biasing the sample substantially. The enhanced angular resolution provided by GH 
provides a better base to detect positional changes in order to study the kinematics of the OB association.

\item {\bf MER}LIN {\bf G}alaxy {\bf E}volution Survey (PI's: T.~Muxlow, I.~Smail, I.~McHardy):\\
{\bf -} L-band, C-band, single pointings and field mapping/mosaicing, (Tier 0 and Tier 1) 918~hours.\\
{\bf -} The science goals rely on reliably measuring the structural information of the radio sources and to 
disentangle extended star-formation from compact AGN. The imaging sensitivity is crucial to this project therefore 
the LT needs to be considered. In addition, the increased angular resolution provided by GH 
will help to better explore the nuclear region at linear scales of 660~pc (L-band) and 140~pc (C-band) respectively (assuming redshift 1).

\item {\bf e}-MERLIN {\bf P}ulsar {\bf I}nterferometry Project (PI: W.~Vlemmings, B.~Stappers):\\
{\bf -} L-band, C-band, targeted imaging, visibility fitting, astrometry, multiple epoch, 160~hours.\\
{\bf -} The LT is considered for special cases, there are FoV restrictions in L-band due to in-beam
  calibrators. The basic observable for astrometry is the position on the sky and with the enhanced angular resolution
 provided by GH such measurements will have an enhanced accuracy of a factor of 2.

\item Feedback Process in Massive Star Formation (PI: M.~Hoare, W.~Vlemmings):\\
{\bf -} C-band; targeted imaging, spectral line imaging of methanol- and hydroxyl, full polarisation, 450~hours.\\
{\bf -} The LT is crucial for this project in order to test various star formation models. The enhanced angular 
  resolution due to GH provide better constrains on how to relate the radio outflow to the 
spectral lines and therefore will provide important clues on the origin of the spectral line emission.

\item Gravitational Lensing and galaxy evolution with \emer\ (PI: N.~Jackson, S.~Serjeant):\\
  {\bf -} L-band , C-band, targeted imaging, 228~hours.\\
  {\bf -} Central images of lenses probe the central (10-50~pc) part
  of the potential well of lens galaxies at significant cosmological
  distance. GH baselines will add sensitivity on spatial scales
  well-matched to the problem, allowing detection of fainter central
  images; the LT is important in adding sensitivity. Inclusion of
  GH/LT will also be important in high-fidelity mapping both of
  existing radio lenses and new lenses discovered with Herschel ATLAS,
  which is also part of the project. Many of these lenses are small
  (<1 arcsecond), and at low declination, and in both cases GH
  baselines are a major advantage.
 
\item {\bf L}egacy {\bf e}-{\bf M}ERLIN {\bf M}ulti-Band {\bf I}maging of {\bf N}earby {\bf G}alaxie{\bf s} (PI: R.~Beswick, I.~McHardy):\\
  {\bf -} L-band, C-band, targeted imaging, line data mining considered, 810~hours.\\
  {\bf -} The aim is to map out radio star-formation indicators in
  nearby galaxies, such as HII regions and supernova remnants (SNR),
  and to carry out a census of low luminosity AGN. Without the LT the
  investigation is biased to the nearest and brightest sources making
  the survey neither complete nor representative. The enhanced angular
  resolution provided by GH will help to map out sub-structures at pc
  scales to better constrain the nature of such indicators and will
  help distinguish unresolved AGN from starburst activity.

\item {\bf L}uminous {\bf I}nfra-{\bf R}ed {\bf G}alaxy {\bf I}nventory (PI: J.~Conway, M. Perez-Torres):\\
{\bf -} L-band, C-band, targeted imaging, line imaging, multiple epoch, full polarisation, 353~hours.\\
{\bf -} The LT is considered necessary to match the sensitivity in polarisation and to detect a substantial rate of RSNe in
 the most extreme starburst environments. Due to the compactness of the RSNe the enhanced angular resolution by GH will
provide a better constraint on the nature of the detected radio emission.

\item Morphology and Time Evolution of Thermal Jets Associated with Low Mass Young Stars (PI: L.~Rodriguez):\\
{\bf -} C-band, targeted imaging, multiple epoch, 180~hours.\\
{\bf -} The LT is needed to detect the hollow-core and the optically thin outer regions of the stellar jets and 
hence lead to a fundamental understanding on the disk-jet relation. GH will provide a better estimate on the proper motion of the sources.

\item {\bf P}lanet {\bf E}arth {\bf B}uilding {\bf B}locks - a {\bf L}egacy {\bf e}-MERLIN {\bf S}urvey (PI: J. Greaves):\\
{\bf -} C-Band, targeted imaging, 72~hours.\\
{\bf -} Only with the LT is e-MERLIN able to detect the cm dust emission on a few AU scales for the first time. 
The improved resolution with GH will better constrain the planet-forming zones and allow dust detections of a few Earth-masses.

\item Resolving Key Questions in Extragalactic Jet Physics (PI: R.~Laing):\\
{\bf -} L-band, C-band, targeted imaging, multi epoch, full polarisation, 375~hours.\\
{\bf -} For a sub-sample the LT is considered, without Lovell the observable sub-sample would
more than halve and hence would not be representative. The increase in angular resolution by GH (80~pc L-Band and 16~pc 
for C-band at redshift 0.05) may be too high to investigate the radio sub-structure of the closest sources of the sample, but 
for the other targets the additional resolution will help to better investigate the study of radio structure.
\end{itemize}}

\noindent Of the 11 Legacy projects, 9 require the additional sensitivity
provided by the Lovell Telescope. Apart from the great advantage in
imaging sensitivity there are some observing constraints due to the
field of view (FoV) limitation mentioned in the proposals, which precludes the use of
the Lovell Telescope. In particular, for in-beam calibration the smaller FoV of the
Lovell Telescope with respect to that of the MKII Telescope will reduce the
number of suitable calibrators. In addition, there is a trade off
between the mapping speed and the sensitivity, such that the mosaicing
efficiency might be reduced if Lovell Telescope is included in \emer .

At least 10 out of 11 proposals seem to accrue significant benefit
from the enhanced resolution provided by Goonhilly.  Except for a
sub-program in one proposal the Legacy programmes will greatly benefit
from this factor 2 increase in spatial resolution, in particular
providing additional constraints on source structure, better
positional constraints of radio components for multi-epoch
observations, and improved astrometry. In addition to the above points
many breakthroughs in these and other areas of science will come from
the combination of \emer\ and EVN data. The much improved overlap of
baselines provided by Goonhilly (Figure 2) will be critical for
reliably achieving this combination, particularly for line
observations where multi-frequency synthesis cannot be used to improve
the UV coverage. 

Finally, it should be noted that the existence of Goonhilly and the
opportunity to access southern fields (below Dec~30$\,^{\circ}$) in \emer\ may
mean that several of the Legacy programmes may wish to re-asses their
selection of targets or survey locations to benefit from overlap with
the ESO facilities such as ALMA or VISTA.

\section{Conclusions}

\noindent The addition of an antenna at Goonhilly to the \emer\ array
provides a near-doubling of the spatial
resolution and an improved UV-plane coverage. The increased N-S extent
also has strong positive imaging implications for equatorial imaging
with the \emer\ array. The very sensitive 440~km baseline that would be formed
between Goonhilly and the Lovell Telescope would result in 
rich enhancements of the \emer\ Legacy surveys, roughly doubling the
overall imaging sensitivity and tripling the long-baseline
sensitivity. Finally, the Goonhilly station provides a unique overlap in baseline
lengths between \emer\ and the EVN, allowing robust combination of
data from the two instruments. Such combined data sets are unique and
open up the ability to study radio morphologies from kilo-parsec up to
parsec scales.

\end{document}